\documentclass[prb,amsfonts,amssymb,floats,twocolumn,aps]{revtex4-1}
\usepackage[utf8]{inputenc}
\usepackage[T1]{fontenc}
\usepackage{amsmath}
\usepackage{float}
\usepackage[caption = false]{subfig}
\usepackage{color}
\usepackage{braket}
\usepackage{bm}
\usepackage{dsfont}
\usepackage{adjustbox}
\usepackage{graphicx}
\usepackage{tikz}
\usepackage{placeins}
\usepackage{bbm}

\newcommand{\iu}{\mathrm{i}} 
\newcommand{\eu}{\mathrm{e}} 
\newcommand{\du}{\mathrm{d}} 
\newcommand{\hc}{\mathrm{h.c.}} 

\newcommand{\Ztwo}{\mathbb{Z}_2}
\newcommand{\Uone}{\mathrm{U(1)}}
\newcommand{\SUtwo}{\mathrm{SU(2)}}

\newcommand{\XY}{{xy}}
\newcommand{\ZZ}{{z}}
\newcommand{\HF}{\mathrm{HF}}

\newcommand{\jxy}{j^\XY}
\newcommand{\jzz}{j^\ZZ}
\newcommand{\tw}{\langle 2 \rangle}
\newcommand{\xyfm}{XY-FM}

\graphicspath{{./}{figs/}}

\begin{document}

\title[]
{
Theory of partial quantum disorder \\ in the stuffed honeycomb Heisenberg antiferromagnet
}
\author{Urban F. P. Seifert}
\affiliation{Institut f\"ur Theoretische Physik,
Technische Universit\"at Dresden, 01062 Dresden, Germany}
\author{Matthias Vojta}
\affiliation{Institut f\"ur Theoretische Physik,
Technische Universit\"at Dresden, 01062 Dresden, Germany}


\date{\today}

\begin{abstract}
Recent numerical results [Gonzalez \textit{et al.}, Phys. Rev. Lett. \textbf{122}, 017201 (2019); Shimada \textit{et al.}, J. Phys. Conf. Ser. {\bf 969}, 012126 (2018)] point to the existence of a partial-disorder ground state for a spin-1/2 antiferromagnet on the stuffed honeycomb lattice, with 2/3 of the local moments ordering in an antiferromagnetic N\'eel pattern, while the remaining 1/3 of the sites display short-range correlations only, akin to a quantum spin liquid.
We derive an effective model for this disordered subsystem, by integrating out fluctuations of the ordered local moments, which yield couplings in a formal $1/S$ expansion, with $S$ being the spin amplitude. The result is an effective triangular-lattice XXZ model, with planar ferromagnetic order for large $S$ and a stripe-ordered Ising ground state for small $S$, the latter being the result of frustrated Ising interactions. Within the semiclassical analysis, the transition point between the two orders is located at $S_c=0.646$, being very close to the relevant case $S=1/2$. Near $S=S_c$ quantum fluctuations tend to destabilize magnetic order. We conjecture that this applies to $S=1/2$, thus explaining the observed partial-disorder state.
\end{abstract}

\pacs{}

\maketitle


\section{Introduction}

Frustration counteracts the tendency of a spin system to order magnetically at low temperatures, and, if strong enough, can lead to quantum disordered ground states devoid of spontaneous symmetry breaking. Such states are conceptually fascinating, a prime example being gapped quantum spin liquids with intrinsic topological order and fractionalized excitations, evoking continued both theoretical and experimental interest.\cite{savary_rop17,ropfru}
A prototypical frustrated spin system is the triangular-lattice antiferromagnet.\cite{anderson73,ms01} Following Anderson's proposal that the spin-$1/2$ Heisenberg model may host a resonating-valence-bond (RVB) state, recent numerical studies confirm that upon amending the nearest-neighbor model with further frustrating interactions, such as second-neighbor exchange, indeed stabilizes a quantum spin liquid.\cite{whiche07,zhuwhi15,sheng15,cherny18}

\begin{figure}[!tb]
\includegraphics[width=\columnwidth,clip]{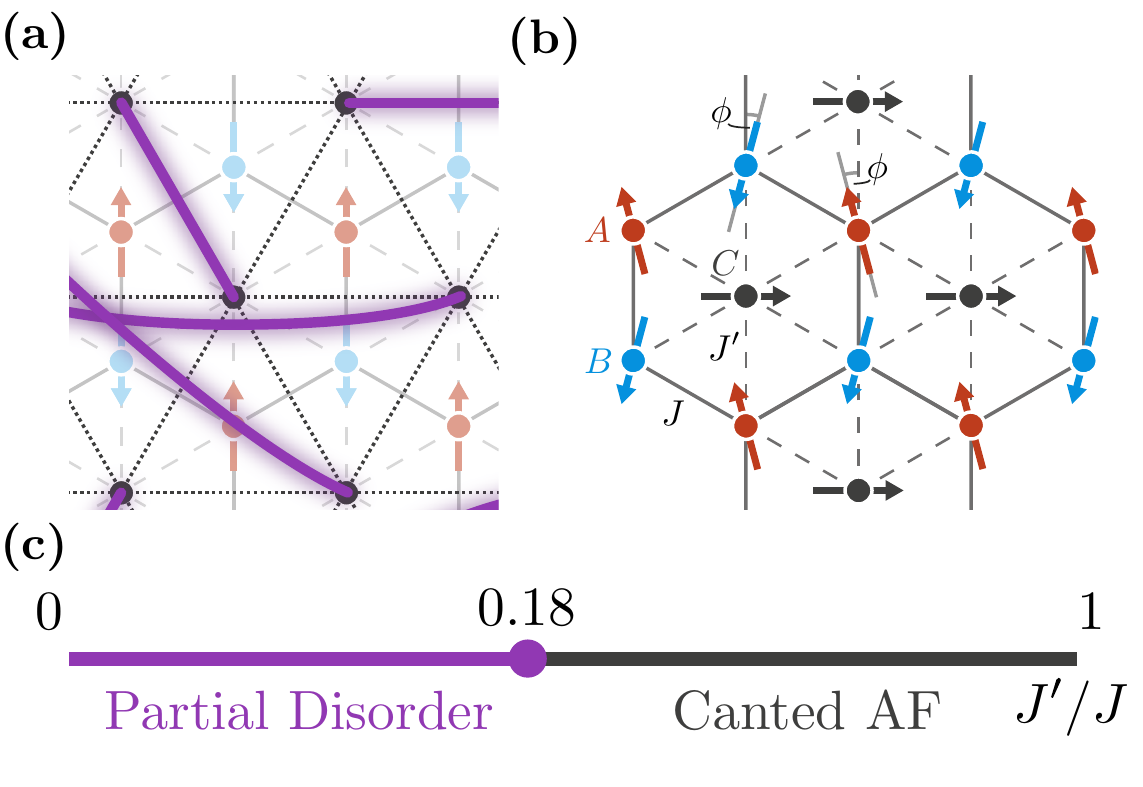}
\caption{
(a) Illustration of the correlated partial-disorder phase,\cite{gonz18} in which the honeycomb spins have collinear N\'eel order and the central spins, located on an effective triangular lattice (dotted), are in a quantum disordered state with short-ranged correlations (possibly RVB-like, illustrated in purple).
(b) Classical ground state of the Heisenberg antiferromagnet on the stuffed honeycomb lattice with $J' \leq J$. The spins on the $A$, $B$ sublattices continuously cant as function of $J'/J$, interpolating between collinear $AB$ order at $J'=0$ and $120^\circ$ $ABC$ order at $J' = J$.
(c) Phase diagram for $S=1/2$ as proposed in Ref.~\onlinecite{gonz18} based on DMRG simulations.
}
\label{fig:model}
\end{figure}

While quantum disorder, non-trivial topology, and fractionalization are typically properties of the whole system, intriguing cases with \textit{coexisting} topological and non-topological components have been discussed in different settings.\cite{flst1,flst2,semoe15} In particular, it has been suggested that frustration may lead to \textit{partially} disordered states, where a disordered subsystem coexists with a magnetically ordered part. While these studies have been mostly concerned with Ising models\cite{meka77,blan84,moto12} or classical systems\cite{diep97} at finite temperature, a recent density-matrix-renormalization-group (DMRG) study by Gonzalez \textit{et al.}\cite{gonz18} has discovered such phenomenology in a SU(2)-invariant quantum spin system.

Ref.~\onlinecite{gonz18} considered a spin-$1/2$ Heisenberg antiferromagnet on the triangular lattice with a $\sqrt{3}\times\sqrt{3}$ superstructure (also dubbed stuffed honeycomb lattice), i.e., a honeycomb lattice with exchange coupling $J$, supplemented by spins in the center of each hexagon which are coupled by $J'$ to the surrounding spins, Fig.~\ref{fig:model}. For a range of finite small couplings $J'/J$, the central spins were found to be decoupled from the N\'eel-ordered honeycomb subsystem.  In this partial-disorder phase, the central spins are short-range correlated, as opposed to previous models of partial disorder (PD), in which the disordered subsystem had a classical extensive degeneracy. Importantly, PD in the stuffed honeycomb lattice is a quantum effect, as the (semi-)classical ground state at any finite $J'<J$ involves ferromagnetic order of the central spins, with the honeycomb spins canting in direction of the ordering axis of the central spins.
The existence of PD physics in the stuffed honeycomb lattice is supported by an exact diagonalization (ED) study which shows a phase with vanishing magnetization at weak $J'$.\cite{shima18}

The goal of this paper is to identify the mechanism which causes the PD state in the stuffed honeycomb-lattice Heisenberg model. The fact that PD exists in the regime of small $J'$ allows us to attack the problem perturbatively: We integrate out the degrees of freedom of the honeycomb subsystem, assuming collinear N\'eel order of the latter. We work to second order in $J'$ and apply a systematic $1/S$ expansion such that our study is formally controlled in the semiclassical (large-$S$) limit.
We obtain an anisotropic effective spin model for the central spins on an emergent triangular lattice. The interactions among the central spins induced by magnons are given by predominantly ferromagnetic transverse ($\XY$) interactions and frustrating Ising interactions for the $\ZZ$-components, with the leading terms arising at different orders in $1/S$:
The leading $\XY$ coupling scales as $S^0$, while the leading Ising interaction scales as $S^{-1}$. As a result, the effective model displays ferromagnetic in-plane order at large $S$, while at small $S$ the Ising interactions become dominant, favoring an out-of-plane stripe-ordered ground state.
With all effective couplings calculated to order $S^{-1}$, we find the transition between the two competing ground states to be located at $S_c=0.646$, being remarkably close to the case of $S=1/2$ considered in the numerical studies. Moreover, the transition is masked by a small window of more complicated (most likely incommensurate) order, and magnetization corrections grow near $S=S_c$. We therefore argue that fluctuations in this frustrated effective model destabilize magnetic long-range order for $S=1/2$, which then leads to a PD phase for the stuffed honeycomb antiferromagnet.

The remainder of the paper is organized as follows:
In Section \ref{sec:model} we introduce the Hamiltonian and review previous numerical results.
Section \ref{sec:derivation} outlines the derivation of the effective spin model, which is discussed in Section \ref{sec:eff_mod}. Section \ref{sec:beyondeff} discusses global properties of the partial-disorder phase beyond the effective model.
A discussion (Section \ref{sec:out}) closes the paper.

\section{Model} \label{sec:model}

We consider local moments on the stuffed honeycomb lattice, consisting of a honeycomb lattice with central spins placed in each hexagon, as depicted in Fig. \ref{fig:model}(b). The situation corresponds to a $\sqrt{3} \times \sqrt{3}$ supermodulation of a triangular lattice.

\subsection{Heisenberg Hamiltonian}

The local moments on the honeycomb lattice (with $A$, $B$ sublattices) are coupled antiferromagnetically with a coupling $J>0$. Each central spin (sublattice $C$) couples with interaction $J'>0$ with its nearest neighbors on the $A,B$ sublattices, such that the Hamiltonian of the full model reads $\mathcal H = \mathcal H_J + \mathcal H_{J'}$ with
\begin{align} \label{eq:ham}
  \mathcal H_J &= J \sum_{\langle i j \rangle} \vec S_{i,A} \cdot \vec S_{j,B} \nonumber \\
  \mathcal H_{J'} &= J' \sum_{\substack{\langle i j \rangle\\s=A,B}} \vec S_{i,s} \cdot \vec S_{j,C}.
\end{align}
At $J'=J$, the model reproduces the Heisenberg antiferromagnet on the isotropic triangular lattice, while for $J'=0$ the system reduces to antiferromagnetically coupled spins on the honeycomb lattice, with the central spins being decoupled.
For the subsequent analysis, it will be convenient to consider a unit cell containing $A,B,C$ sites, with the primitive lattice vectors given by $\vec n_{1,2} = (\pm 1, \sqrt{3})^T/2$.

\subsection{Previous numerical results} \label{sec:prev_num}

The classical ground state of the Eq.~\eqref{eq:ham} on the stuffed honeycomb lattice for $0 \leq J '\leq J$ is depicted in Fig.~\ref{fig:model}(b), with the $A,B$-sublattice spins canting at an angle $\phi = \arcsin(J'/(2J))$ to the direction transverse to the $C$ sublattice. This leads to the familiar $120^\circ$ triangular-lattice order at $J=J'$ and collinear order of the honeycomb spins in the limit $J'/J\to 0$. The classical system thus displays ferrimagnetic behaviour at $0< J' < J$, with the ground state having a non-zero magnetization in the spin direction of the central spins.

The authors of Ref.~\onlinecite{gonz18} focus on the case of small $J'/J$ and find that the classical order persists in linear spin-wave theory (LSWT) even for infinitesimal $J'$, with the $C$ sublattice magnetization $m_C$ taking more than $80 \%$ of its classical value.
Their DMRG results for the spin-$1/2$ system, however, show clear signatures of a PD phase at $0 < J' \leq J_c' \approx 0.18$, with vanishing $m_C=0$ and collinear N\'eel order on the honeycomb subsystem. Notably, in PD the central spins are found to be ferromagnetically correlated, hinting towards the presence of an effective interaction between these spins.
For $J'>J_c'$ a canted state is found, with canting angles closely matching the LSWT result.
At $J_c'$ the central-spin magnetization $m_C$ changes abruptly (as do nearest-neighbor spin correlators). Altogether, this points toward a first-order transition between a PD state and a semiclassical canted state as function of $J'/J$, Fig.~\ref{fig:model}(c).

Furthermore, recent ED results by Shimada \textit{et al.}\cite{shima18} for systems up to $36 \times 36$ sites indicate that the spontaneous magnetization of the semiclassical ferrimagnetic phase for $J' < J$ vanishes non-continuously below some non-zero $J_c'$. This would be consistent with the existence of PD, however the value of the critical $J_c'$ and precise nature of the correlations below $J_c'$ appear to be difficult to obtain in ED due to limited accessible system sizes.

\section{Derivation of an effective model} \label{sec:derivation}

The PD phase is realized for small $J'$, suggesting that the coupling between the honeycomb spins and central spins can be treated perturbatively. In the limit $J' \ll J$ it appears justified to neglect feedback effects of the central spins on the honeycomb spins, such that the natural effective model for the system involves central spins on an emergent triangular lattice, with effective interactions resulting from the coupling to the honeycomb system.

The goal of this section is thus to derive this effective spin model for the central spins by perturbatively integrating out quantum fluctuations around the N\'eel-ordered ground state of the honeycomb spins, which we treat in a $1/S$ expansion. For a detailed derivation we refer the reader to Appendix \ref{sec:deriv_details}.

\begin{figure}[!tb]
\includegraphics[width=\columnwidth,clip]{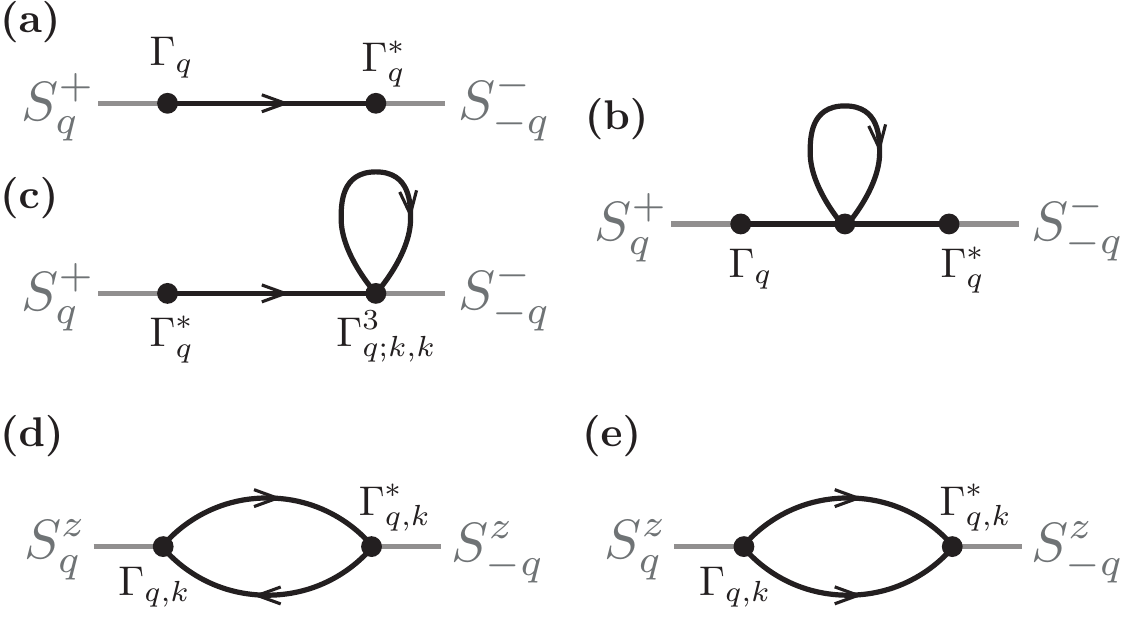}
\caption{
Feynman diagrams for processes contributing to the effective action (symmetry-related processes are not shown). (a) Leading-order contribution $S^0$ to $J^{\XY}$, (b) Hartree-Fock correction to the propagator, of order $1/S$, (c) Correction to magnon-spin vertex at order $1/S$ (d) Particle-hole (vanishes at $T=0$) and (e) particle-particle bubbles mediating a longitudinal interaction at order $1/S$.
}
\label{fig:diags}
\end{figure}

We emphasize that the magnons in the honeycomb system are gapless, and integrating out theses gapless degrees of freedom may in general induce singular interactions. We thus apply a staggered field $h$ in direction of the N\'eel order on the honeycomb sublattices, giving rise to a gap in the magnon dispersion, and take the massless limit $h \to 0$ at the end.

Without loss of generality, we parameterize the semiclassical N\'eel order of the honeycomb system by $\vec S_A = S \hat{z}$ and $\vec S_B = -S \hat{z}$ such that the Hamiltonian with a staggered field reads $\mathcal{H}_{J,h} =\mathcal{H}_J - 3 J S h \sum_i (S_{i,A}^z - S_{i,B}^z)$.
We consider magnon excitations on top of this ordered ground state by representing the $\SUtwo$ spin algebra in terms of Holstein-Primakoff bosons $a,b$, admitting a systematic expansion in $1/S$.
The $1/S$ expansion is formally justified for $S \gg n_{a,b}$, where the bosonic densities $n_a=a^\dagger a$, $n_b=b^\dagger b$ quantify the mean deviation from the classical reference state. This implies that the expansion is applicable (even for $S \lesssim 1$) for magnetically ordered states with small intrinsic fluctuations, such as the collinear state on the honeycomb lattice of interest here.
We expand $\mathcal H_J = \mathcal H_{J,h}^{(0)} +\mathcal H_{J,h}^{(2)}+\mathcal H_{J,h}^{(4)}+\mathcal{O}(1/S)$ where $\mathcal H^{(n)}$ contains $n$ boson operators. The leading order $\mathcal H_{J,h}^{(0)} \sim S^2$ represents the classical ground-state energy. The bilinear piece corresponds to LSWT and can be written as
\begin{equation} \label{eq:Hjh_lswt}
  \mathcal{H}^{(2)}_{J,h} = J S \sum_{q} \left[ f(q) a_q b_{-q} + \mathrm{h.c.} + 3(1+h)(a_q^\dagger a_q + b_q^\dagger b_q) \right],
\end{equation}
where $f(\vec q) = 1+ \eu^{\iu \vec q \cdot \vec n_1} + \eu^{\iu \vec q \cdot \vec n_2}$ with $\vec n_{1,2}$  defined as above.
$\mathcal{H}^{(2)}_J$ can be diagonalized by means of a Bogoliubov transformation, yielding two magnon modes (having neglected a global energy shift and employing inversion symmetry $\epsilon(q)=\epsilon(-q)$)
\begin{equation} \label{eq:Hjh_lswt_diag}
  \mathcal{H}^{(2)}_{J,h} = \sum_q \epsilon(q) (\alpha_q^\dagger \alpha_q + \beta_{q}^\dagger \beta_{q}),
\end{equation}
where $\epsilon(q) = J S \omega(q)$ with the dimensionless dispersion $\omega(q) = \sqrt{9 (1+h)^2 - |f(q)|^2}$.
The quartic part of the Hamiltonian corresponds to magnon-magnon interactions,
\begin{align} \label{eq:quartic}
  \mathcal{H}^{(4)}_{J,h} = -\frac{J}{4} \sum_{i,\delta} \left[ a_i^\dagger a_i a_i b_{i+\delta} + a_i b_{i+\delta}^\dagger b_{i+\delta} b_{i+\delta} \right. \nonumber \\ \left. + 2 a_i^\dagger a_i b^\dagger_{i+\delta} b_{i+\delta} +\hc \right];
\end{align}
we note that cubic boson terms are absent in the present collinear case.
The quartic terms give rise to self-energy corrections to the magnon Green's function. In general, the self-energy corrections can be split into a static Hartree-Fock contribution $\Sigma^\HF$ arising from ``balloon'' diagrams as shown in Fig.~\ref{fig:diags}(b)  and a frequency-dependent contribution represented by ``sunset'' diagrams, with the latter scaling as $1/S$.\cite{zhitom09,fn:dimless} An explicit expression for $\Sigma^\HF$ is given in Appendix \ref{sec:deriv_details}.

The interaction vertices for the coupling of the magnons to the central spins $\vec S_i$ (omitting the label $C$ for brevity) can be determined by expanding $\mathcal{H}_{J'}$ in $1/S$, yielding up to third order
\begin{widetext}
\begin{align} \label{eq:h_vertices}
  \mathcal{H}_{J'}  = J' \sum_q \left[ \left( \sqrt{\frac{S}{2}} \Gamma^a_q a_q + \frac{1}{4\sqrt{2S}} \sum_{k,p} \Gamma^{3a}_{q;k,p} a_{k+p-q}^\dagger a_k a_p + \sqrt{\frac{S}{2}}\Gamma^b_q b_{-q}^\dagger + \frac{1}{4\sqrt{2S}}\sum_{k,p} \Gamma^{3b}_{q;k,p} b_{-k}^\dagger b_{-p}^\dagger b_{-(k+p-q)} \right) S^-_{-q}  + \mathrm{h.c.} \right. \nonumber \\
  \left. + \sum_k \left( \Gamma_{q,k}^{aa} a_{k+q}^\dagger a_k + \Gamma_{q,k}^{bb} b_{k+q}^\dagger b_k \right) S^z_{q} \right],
\end{align}
\end{widetext}
where we have made the $S$ scaling for the respective vertices explicit (for definitions of the various vertices we refer the reader to Appendix \ref{sec:deriv_details}).
Note that Eq.~\eqref{eq:h_vertices} does not contain a term linear in $S$, which corresponds to the fact that mean fields on the central spins vanish -- apparently a prerequisite to obtain a PD phase.

We are now in position to integrate out the bosonic modes, yielding an effective model for the central spins which are arranged on a triangular lattice. At order $(J')^2$ and next-to-leading order in $1/S$ the effective action takes the form
\begin{align} \label{eq:seff}
  \mathcal S_\mathrm{eff} = \frac{\left(J'\right)^2}{J} \int \du \tau \int \du \tau' &\sum_q S^+_q(\tau) \jxy(q,\tau-\tau') S^-_{-q}(\tau') \nonumber
  \\ + &S^z_q(\tau) \jzz(q,\tau-\tau') S^z_{-q}(\tau')
\end{align}
with dimensionless couplings $j$.
The relevant processes contributing to this order are shown in Fig.~\ref{fig:diags}:
The coupling $\jxy$ as a result of transverse magnon modes is given by the magnon propagator and subleading Hartree-Fock corrections for both the propagator and vertex, while the longitudinal $\jzz$ results from magnon bubble diagrams.

\begin{figure}[!tb]
\includegraphics[width=.9\columnwidth,clip]{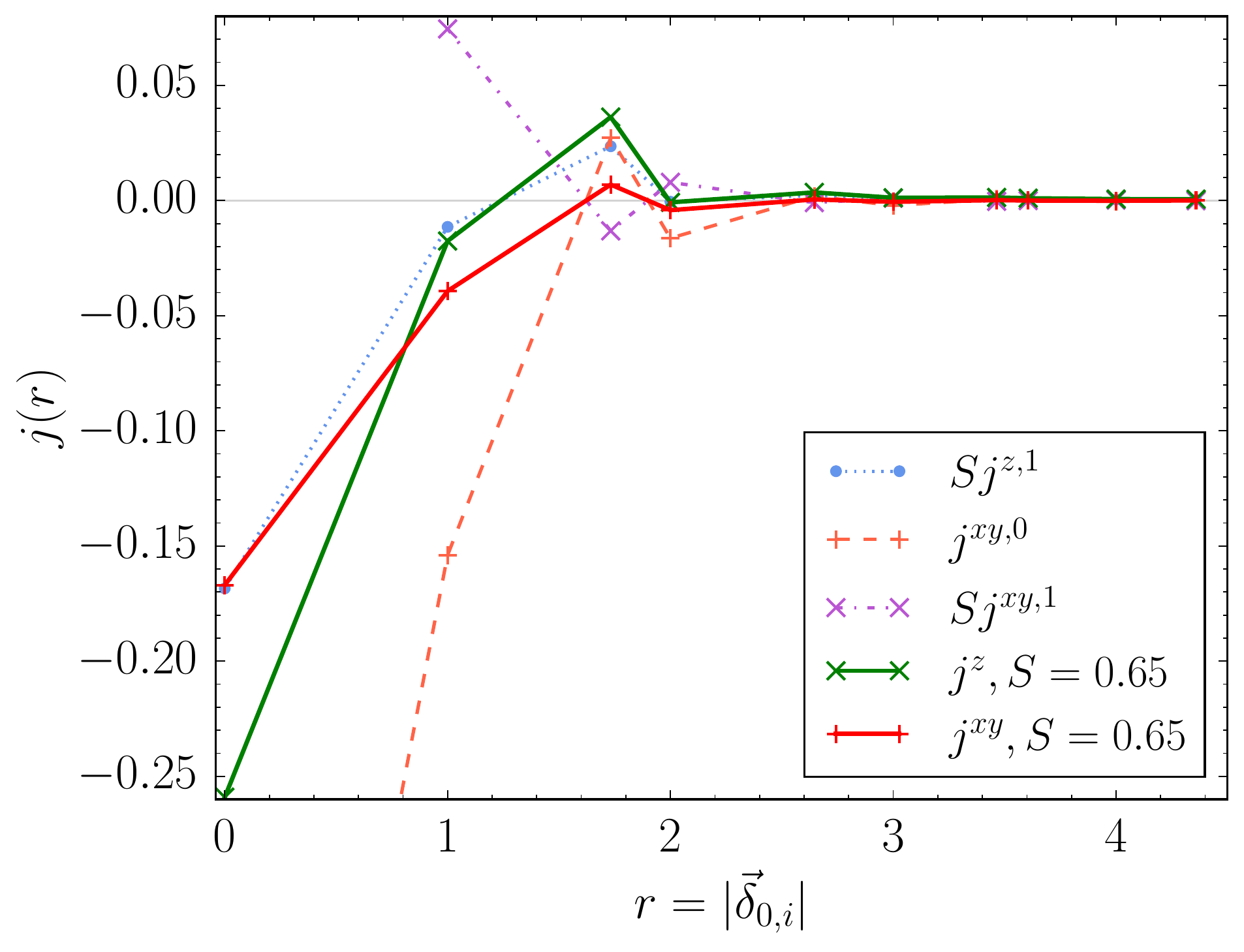}
\caption{
  Dimensionless couplings in the effective triangular-lattice XXZ model as a function of the real-space distance to the $i$-th nearest neighbor. Shown are the coefficients of the $1/S$ expansion $j^{\XY,0}$, $Sj^{\XY,1}$, $Sj^{\ZZ,1}$ as well as the  couplings evaluated for $S=0.65$, i.e., close to $S=S_c$.
}
\label{fig:overview_coups}
\end{figure}

 The effective model manifestly breaks $\SUtwo$ spin rotation symmetry, and is invariant only under a $\Uone$ symmetry of rotations about the $z$ axis. Physically, this is due to the background spin order of the honeycomb subsystem, which itself spontaneously breaks $\SUtwo$ spin rotation symmetry at $T=0$ (see Sec.~\ref{subsec:exps} for a discussion of the case $T>0$), i.e., the effective model simply inherits the symmetry of the full system. Considering a generic large-$S$ expansion around a classically ordered state (i.e. in a symmetry-broken phase), it is easily seen that the $\SUtwo$ symmetry is indeed broken at every order in $1/S$.

The time dependence of the couplings in Eq.~\eqref{eq:seff} is a consequence of retardation effects. In order to obtain a ``static'' spin Hamiltonian $\mathcal H_\mathrm{eff}$ for the central spins, we employ the instantaneous approximation $j(q,\tau-\tau') \to j(q) \delta(\tau-\tau')$, where $j(q)$ is given by the $\omega \to 0$ limit.
This approximation is justified as long as there is a separation of energy scales between the fast (high-energy) magnon modes at an energy scale $J$ and the low-energy central spins at a scale $J' \ll J$.

We thus find the static transversal coupling to be given by
\begin{equation} \label{eq:jxy_q}
  \jxy (q) = - \frac{1}{2 \omega_q} \sum_{\delta=\alpha,\beta} \left[ |\Gamma^\delta |^2 \left(1  - \frac{\Sigma^\HF_q}{S \omega_q} \right) + \frac{\Re \left[\Gamma^\delta \Gamma^{3 \delta \ast}_q \right]}{4 S}  \right],
\end{equation}
and the longitudinal coupling
\begin{equation} \label{eq:jzz_q}
  \jzz (q) = - \frac{1}{2 S} \sum_k \left[ \frac{|\Gamma^{\alpha \beta}_{q,k}|^2 + |\Gamma^{\beta \alpha}_{q,k}|^2}{\omega(k) + \omega(q+k)} \right],
\end{equation}
with the explicit form of the vertices $\Gamma$ detailed in Appendix \ref{sec:deriv_details}.
The couplings given above implicitly depend (through the dispersion and the Bogoliubov factors appearing in the vertices) on the staggered field strength $h$.
In order to extrapolate $h \to 0$ and to identify the most important spin interactions, it is convenient to work in real space. Fourier-transforming thus leads to the effective Hamiltonian
\begin{equation} \label{eq:heff}
  \mathcal{H}_\mathrm{eff} = \frac{(J')^2}{J} \sum_{n=0}^\mathcal{N} \sum_{\langle ij \rangle_n} \left[ \jxy_{ij} \left(S^x_i S^x_j + S^y_i S^y_j\right) + \jzz_{ij} S^z_i S^z_j \right],
\end{equation}
where $\langle ij \rangle_n$ denotes a bond between $n$-th nearest neighbors on the triangular lattice of central spins. Working to order $S^{-1}$, the transverse coupling $\jxy_{ij} = j^{\XY,0} + j^{\XY,1}$ contains both a leading $j^{\XY,0} \sim S^0$ and subleading $j^{\XY,1} \sim S^{-1}$ contribution, while the longitudinal coupling scales as $\jzz_{ij} = j_{ij}^{\ZZ,1} \sim S^{-1}$.

We truncate the generated (long-ranged) interactions after the $\mathcal N$-th nearest neighbor, finding that the properties of the effective model, discussed in Sec.~\ref{sec:eff_mod}, do not significantly depend on the truncation after the most dominant interactions are taken into account.
The $h\to 0$ limit is taken by first evaluating the expressions in Eqs. \eqref{eq:jxy_q} and \eqref{eq:jzz_q} for fixed $h$, and then fitting the results to a leading-order scaling form in $h$ obtained by a continuum approximation.
For details on the extrapolation, we refer the reader to Appendix \ref{sec:extrapol}. The obtained values of the couplings are shown in Table~\ref{tab:effcoups}.

We note that the above scheme naturally yields on-site spin couplings ($n=0$), corresponding to (retarded) self-interactions of the spins mediated by magnons. In the instantaneous approximation, the on-site couplings $j_{ii}^{\XY} \neq j_{ii}^{\ZZ}$ combine to a single-ion anisotropy for any $S > 1/2$, while simply yielding a global energy shift at $S=1/2$. In order to allow for a consistent large-$S$ study, we henceforth include the single-ion anisotropy in the analysis of the effective model (i.e. in classical iterative minimization and linear spin-wave theory).\cite{fn:nosingleion}

\begin{table}[htbp]
  \centering
    \caption{Effective couplings obtained for model \eqref{eq:heff} for the $n$-th nearest neighbors. Note that $n=0$ corresponds to an anisotropic on-site coupling, yielding a single-ion anisotropy for any $S>1/2$.}
    \begin{tabular}{c|c|c|c}
    \hline
    $n$ & $j^{\XY,0}$ & $S j^{\XY,1}$ & $ S \jzz$ \\
    \hline\hline
    0 & $-0.653991$ & $0.3165606$ & $-0.1684154$ \\
    \hline
    1 & $-0.153987$ & $0.0745365$ & $-0.0114266$ \\
    2 & $0.027187$ & $-0.0131602$ & $0.0235538$ \\
    3 & $-0.016334$ & $0.0079060$ & $-0.0005096$ \\
    4 & $0.001881$ & $-0.0009108$ & $0.0022996$ \\
    5 & $-0.002153$ & $0.0010418$ & $0.0007775$ \\
    6 & $0.000684$ & $-0.0003309$ & $0.0008679$ \\
    7 & $0.000021$ & $-0.0000099$ & $0.0006570$ \\
    8 & $-0.000385$ & $0.0001868$ & $0.0004147$ \\
    9 & $0.000150$ & $-0.0000716$ & $0.0003788$ \\
    \hline
    \end{tabular}%
  \label{tab:effcoups}%
\end{table}

\section{Analysis of the effective model} \label{sec:eff_mod}

The anisotropic effective model obtained in the previous section contains an $\XY$ interaction which has a dominant ferromagnetic coupling on nearest neighbors to order $S^0$. As visible from Table~\ref{tab:effcoups}, this coupling gets significantly reduced at order $S^{-1}$ by self-energy and vertex corrections. The most important $\ZZ$-Ising coupling is an antiferromagnetic \textit{second-neighbor} interaction.

\subsection{Qualitative discussion of dominant processes} \label{sec:qual_discuss}

\begin{figure}[!tb]
\includegraphics[width=\columnwidth,clip]{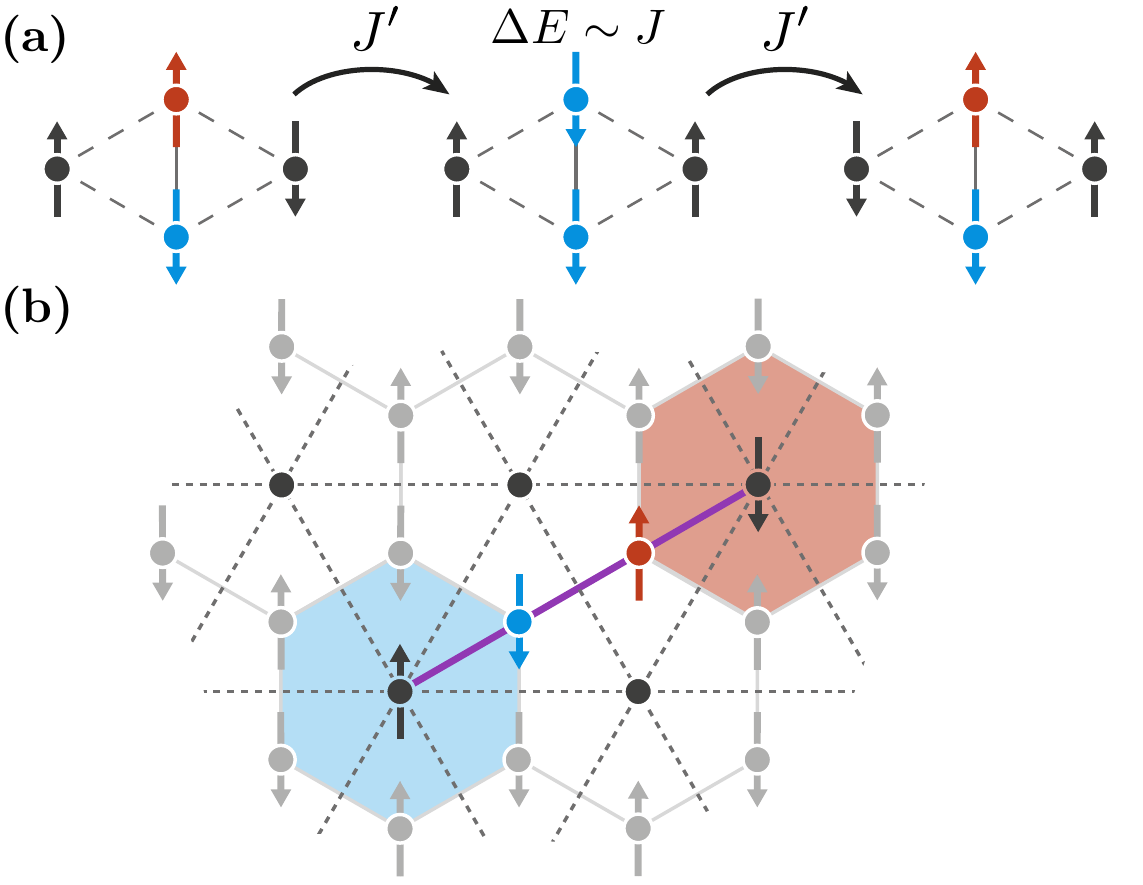}
\caption{
Processes giving rise to the dominant $\XY$ and $\ZZ$ interactions in the effective model.
(a) Exchange process for antiferromagnetic initial and final configurations in second-order peturbation theory, involving a honeycomb-spin flip with energy cost $J$. This process is not realized for parallel configurations and thus gives rise to a ferromagnetic transversal interaction at order $S^0$.
(b) The presence of two honeycomb-spin flips (representative for a coherent two-magnon excitation, pictured by colorized spins) on a nearest-neighbor bonds gives rise to a antiparallel fields on central spins in next-nearest hexagons,\cite{fn:no_jprime} yielding an effective antiferromagnetic Ising $\ZZ$ interaction (indicated in purple) for second neighbors on the effective triangular lattice (dashed) at order $S^{-1}$.
}
\label{fig:pert_theory_fig}
\end{figure}

We quickly discuss the physical picture behind the most important effective interactions. The nearest-neighbor $\XY$ (transverse) interaction is pictorially shown in Fig.~\ref{fig:pert_theory_fig}(a). Two  nearest-neighbor central spins can exchange their spin orientation with an intermediate honeycomb-spin flip (which, loosely speaking, can be understood as a one-magnon excitation). As the energy cost of the intermediate state is $\Delta E \sim J$, second-order perturbation theory leads to matrix elements of the form
\begin{equation}
  \langle \uparrow \downarrow | \mathcal{H}_\mathrm{eff} |\downarrow \uparrow \rangle \sim - \frac{(J')^2}{J},
\end{equation}
corresponding precisely to the scaling of the effective action in Eq.~\eqref{eq:seff}. Note that this exchange process constitutes the leading-order contribution to $\jxy$. It can only be realized for antiparallel nearest-neighbor spins in the initial and final configurations, thus leading to an effective ferromagnetic $\XY$ interaction, as also observed in the effective model.
We note that a dominant ferromagnetic nearest-neighbor coupling in the effective model is consistent with the observed ferromagnetic spin correlation for PD in the DMRG data.\cite{gonz18}

The dominant contribution of the (longitudinal) $\ZZ$-Ising interaction is antiferromagnetic on next-nearest neighbors. The computation of this coupling (cf. Sec.~\ref{sec:derivation}) involves the evaluation of a magnon-bubble diagram, as displayed in Fig.~\ref{fig:diags}, corresponding to the coherent two-magnon excitation.
This process can be illustrated by considering flipping two neighboring spins on the N\'eel-ordered honeycomb lattice (we emphasize that the spin-flip is \emph{not} equivalent to a magnon excitation, which is rather a coherent superposition of the spin flips due to the Bogoliubov transformation), as shown in Fig.~\ref{fig:pert_theory_fig}(b).
This double spin flip induces antiparallel fields along the spin quantization axis on next-nearest hexagons, therefore favoring an antiparallel alignment of the central spins in the respective two hexagons. This process hence induces an antiferromagnetic Ising interaction of next-nearest neighbors on the effective triangular lattice.
It appears likely that this interaction is further enhanced when taking into account magnon-magnon interactions, because of nearest-neighbor magnon attraction. However, this only contributes to $\jzz$ at order $S^{-2}$ and is not included in the present calculation.

\subsection{Classical ground states}

\begin{figure}[!tb]
\includegraphics[width=.85\columnwidth,clip]{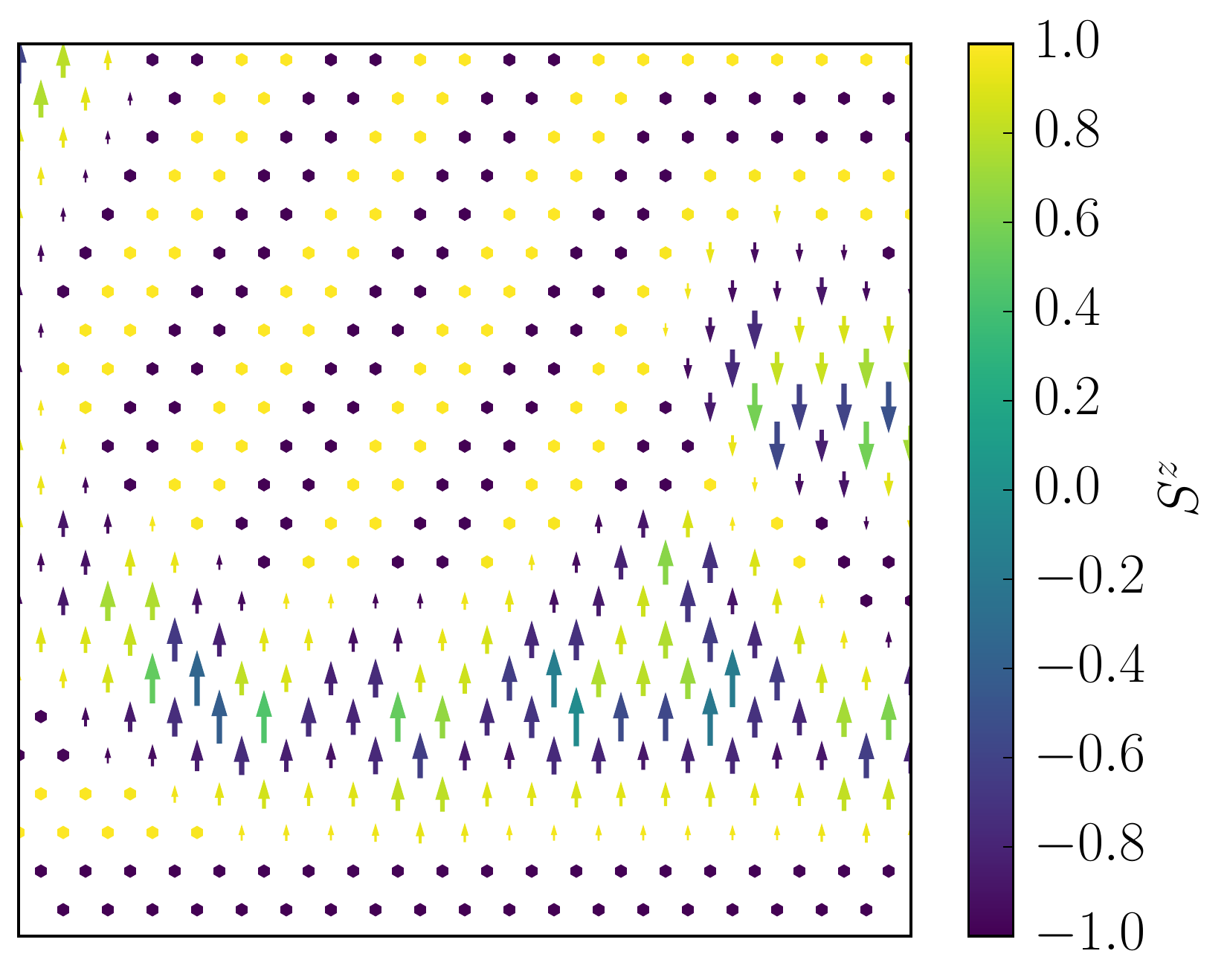}
\caption{
  Metastable configuration obtained through iterative minimization of the classical Hamiltonian \ref{eq:heff} for $S=0.64$ on a $36 \times 36$-site cluster (shown: zoom), illustrating the competition between in-plane ferromagnetic and out-of-plane stripe order. The arrows represent the projection of the spins in the $\XY$ plane, with the color indicating the strength of the $\ZZ$-component. Clearly visible are domains of $\tw$-ordering. The corresponding domain walls are accompanied by in-plane ferromagnetic ordering.
}
\label{fig:config}
\end{figure}

Since the strengths of $\XY$ and $\ZZ$ interactions display a different scaling as function of $S$, we expect the nature of the ground state to change as function of $S$. We hence inspect the \textit{family} of effective models parameterized by $S$: For each fixed $S$ we have a specific triangular-lattice XXZ model, which we again analyze in a formal $1/S$ expansion.

In this subsection, we start by considering the classical ground states. Noting that at $S \gg 1$ only the $\XY$ interaction in \eqref{eq:heff} is of relevance (and subleading corrections are small), while at sufficiently small $S$ the $\ZZ$ terms are dominant, it is legitimate to consider the two interactions separately.
First, we determine the order favoured by the $\XY$ interaction by finding the wavevector $Q$ corresponding to minima of $\jxy(q)$ according the Luttinger-Tisza method. One finds a unique minimum at $Q=0$ for any $S$, corresponding to ferromagnetic order with spins lying in the $\XY$-plane (subsequently dubbed ``\xyfm'').

The Luttinger-Tisza method cannot straightforwardly applied to the case of pure Ising interactions. Instead, we approach the problem of finding the ground state of The Ising part of the Hamiltonian by iteratively minimizing the energy on finite lattices of various sizes. We find that the system orders in a two-up-two-down stripe phase, commonly dubbed $\tw$,\cite{fiselke80} which spontaneously breaks the $C_3$ triangular lattice rotation symmetry, corresponding to the ordering wave vector $Q_{\tw}=(0, \pi/\sqrt{3})^T$ (and $C_3$-symmetry related).
The phase diagram of the Ising model with competing (i.e. ferro-antiferro on nearest and next-nearest neighbors) interactions on the triangular lattice has been explored extensively by Kaburagi and Kanamori, confirming $\tw$ as the ground state in the parameter range relevant to our model.\cite{kaka74,kaka78}
It has been noted that in general competing short-range ferromagnetic and longer-range dominantly antiferromagnetic interactions often favour striped phases,\cite{lieb11,giu16} and can lead to rich phase diagrams with exotic critical properties.\cite{fiselke80,mumaku95,jinsensan12}

Returning to the combined XXZ model for a given $S$ in Eq. \eqref{eq:heff}, we study the ($S$-dependent) competition of transverse and longitudinal interaction terms at the classical level. Numerical results on various system sizes show that above $S_c \simeq 0.6462$, the classical ground state is given by \xyfm{}, while for smaller $S$ the lowest-energy configuration is given by $\tw$, see also Fig.~\ref{fig:config}.
This appears to be consistent with the qualitative discussion above in which we noted that at large $S$ the ferromagnetic $\XY$ interaction is dominant, while the $\jzz$ coupling becomes important compared to $\jxy$ for sufficiently small $S \lesssim 1$.

We note that there is a small window around $S_c$ in which both \xyfm{} and $\tw$ become unstable in favor of an incommensurate phase (dubbed ``IC''). To map out the extent of this window, it is more convenient to use LSWT, as described in the following section.

\subsection{Magnetization corrections in linear spin-wave theory}

\begin{figure}[!tb]
\includegraphics[width=.9\columnwidth,clip]{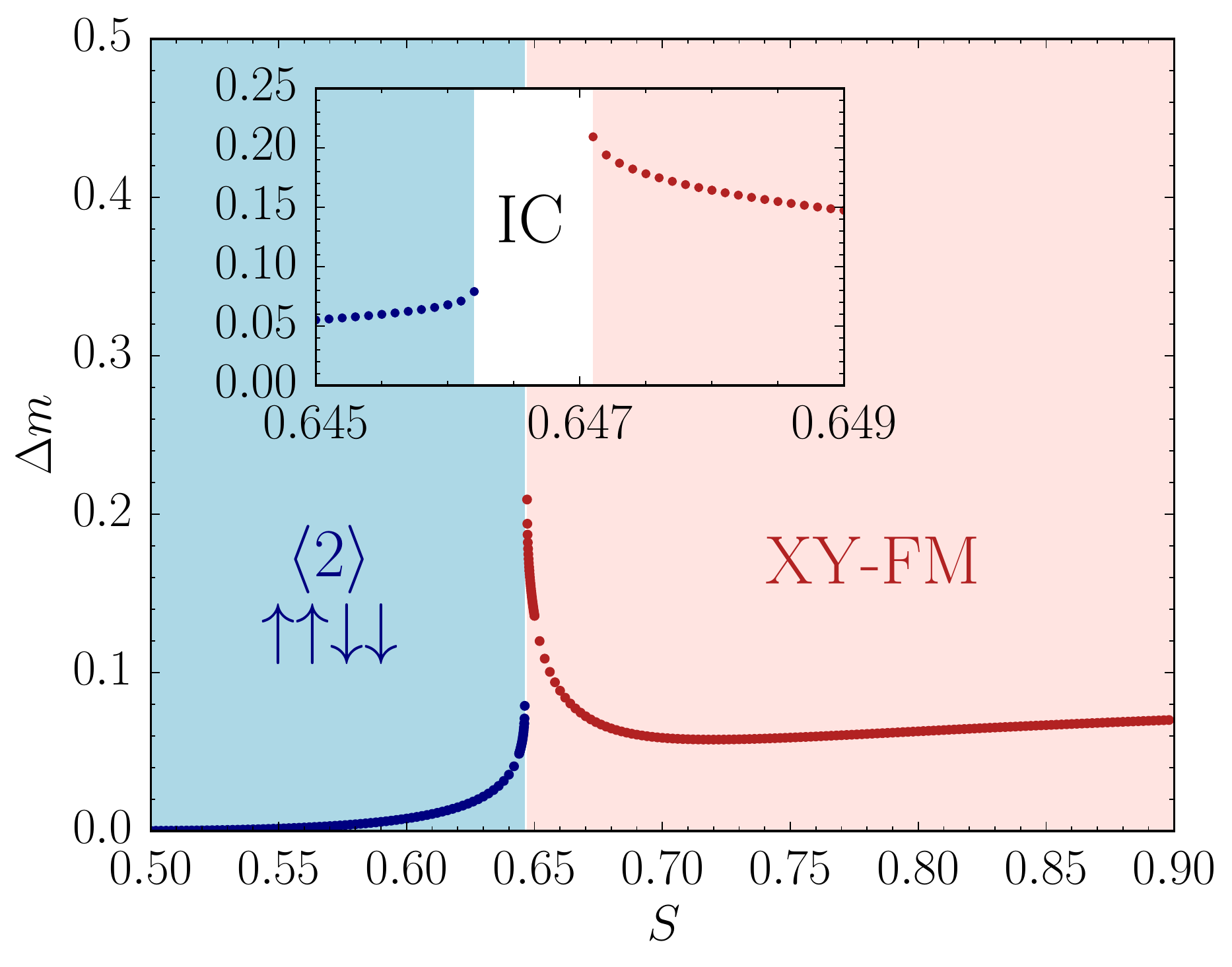}
\caption{
  Classical ground states and corrections to the (staggered) magnetization obtained in LSWT for the effective model \eqref{eq:heff}. In the white region marked IC, LSWT for both \xyfm{} and $\tw$ break down, and the classical ground state is given by an incommensurate modulated stripe phase.
}
\label{fig:LSWT_delta_m}
\end{figure}

The fact that the transition between ferromagnetic and stripe order occurs at $S_c$, very close to the case of $S=1/2$ considered in the numerical studies, indicates that the physics of PD is closely linked to the appearance of this transition. It is also clear that the competition of \xyfm{} and $\tw$ at small $S$ (and thus PD) is an inherently quantum effect, driven by spin fluctuations of the honeycomb lattice. To further explore the competition between the $\XY$ and $\ZZ$ interactions in $\mathcal{H}_\mathrm{eff}$ and the resulting quantum fluctuations, we consider magnon excitations on top of either ferromagnetic or the striped ground state of the $S$-dependent effective model \eqref{eq:heff} in LSWT.

Expanding about the ferromagnetic state, we find that the spin-wave dispersion becomes imaginary for $S<0.6471$. The dispersion closes at the (incommensurate) wave vector $Q_\mathrm{IC}=(0,1.959)^T$ (and $C_3$-symmetry related).
Complementary, expanding around the classical $\tw$ reference state, the spin-wave dispersion becomes imaginary for $S>0.6462$ at a very small $Q$.
The finite window $0.6462 < S < 0.6471$ in which LSWT for both \xyfm{} and $\tw$ breaks down hints at an additional phase, with a possible incommensurate ordering as indicated by the gap closing in LSWT in the \xyfm{}.
Iterative minimization in the IC window, e.g. at $S=0.6469$ indeed yields configurations in which the spins align ferromagnetically in the $\XY$-plane and have a (incommensurately) modulated $\ZZ$-component. Inspecting the static structure factor, we find the ordering wave vector to be given by $Q_\mathrm{IC}$ (to our numerical accuracy). The ordering pattern appears to be related to $\tw$, as also seen from the fact that $Q_\mathrm{IC}$ is close to $Q_{\tw}$.
We have checked on finite-size lattices that the classical energy of the incommensurate configuration is indeed lower than the competing commensurate reference states.
This incommensurate phase thus masks a critical value $S_c$ at which the energies \xyfm{} and $\tw$ are degenerate, as we demonstrate employing a toy model in Appendix \ref{sec:toy_stripe}.
The phase diagram resulting from this discussion is in Fig.~\ref{fig:LSWT_delta_m}.

In addition, we have computed the magnetization corrections $\Delta m$ in LSWT for the two respective ground states, these are also shown in Fig.~\ref{fig:LSWT_delta_m}.
We note a strong increase of the magnetization correction as the spin-wave theory breaks down when tuning towards IC from both sides. Therefore we consider it likely that long-range order disappears completely near $S_c$ once quantum fluctuations are fully taken into account.

Given that the derivation of the effective model is based on a $1/S$ expansion (and involves further approximations, cf. Sec.~\ref{sec:derivation}), we conjecture that the true $S_c$ is close to the physical value $S=1/2$, and the central spins are in a quantum disordered state for $S=1/2$. This then yields the PD state observed in numerical studies.\cite{gonz18,shima18}
Clearly, higher orders in $1/S$ as well as higher orders in $J'$, the latter generating multi-spin exchange interactions which themselves tend to destroy long-range order,\cite{motru04} may be important for a fully quantitative understanding.

It is worth emphasizing that the theme of supplementing a frustrated Ising model by (ferromagnetic) transverse $\XY$ interactions, yielding strong quantum fluctuations, is also key for stabilizing a $\Ztwo$ spin liquid in the easy-axis kagome-lattice spin model studied by Balents, Fisher, and Girvin,\cite{bfg02} and the $\Uone$ spin liquid in the pyrochlore lattice.

\section{Correlated partial disorder beyond the effective model}
\label{sec:beyondeff}

In this section we comment on features of the putative correlated PD phase beyond our effective model.

\subsection{Energetic competition between PD and (semi-)classical canting}

As visible from the phase diagram in Fig.~\ref{fig:model}(c), the PD phase competes energetically with semi-classical canted state. We thus discuss the energy contributions of the two competing phases as a function of $J'/J$.

In the classical ground state of the model (as discussed in Sec.~\ref{sec:model}), the honeycomb spins cant at an angle $\phi = \arcsin(J'/(2J))$ with the central spins pointing in a direction transverse to the N\'eel order. For $J' \ll J$ this yields a canting angle $\phi \propto J'/J$. Considering that the interaction between honeycomb and central spins is of the form $J' \vec S_{A/B} \cdot \vec S_{C}$ and using that $\vec S_{A/B} \cdot \vec S_{C} \sim \phi \sim J'/J$ for small canting angles, we thus find that the mean-field energy of central spins in the canted phase scales as $J'^2/J$. Further we have $\vec S_{A} \cdot \vec S_B \sim 1-\phi^2/2$. Taken together, we see that the energy gain of the canted state relative to the decoupled (collinear) state at $J'=0$ scales as $J'^2/J$.

The energetics of the PD phase is determined by the effective model for the central spins, while mean-field energies between the two subsystems vanish. The effective model was derived in second-order perturbation theory, see Eq.~\eqref{eq:seff}. It is thus clear that the energy gain of the PD phase, relative to the decoupled state at $J'=0$, also scales as $J'^2/J$.

With the two phases having the same energetic $J'$ scaling, no further qualitative arguments can be made as to which phase has a lower energy. We conclude that -- if the PD phase has lower energy than the canted phase for small $J'/J$, as indicated by the numerics\cite{gonz18,shima18} -- the first-order transition at $J'_c$ will occur at $J'_c/J$ of order unity, where contributions in higher order in $J'/J$ become important.

\subsection{Topological properties}

In PD, the central spins are in a correlated quantum disordered phase. Although the precise nature of the phase cannot be obtained on the level of our analysis, it appears likely that this disordered state of the central spin subsystem by itself possesses topological order. Note that this implies that the notion of topological order is expected to be applicable to the entire system, in the sense of the existence of superselection sectors.

We mention similarities to two systems in which a subsystem is in topologically state. First, a topological spin-glass phase has been proposed in diluted spin ice, in which defect-induced ``ghost spins'' eventually freeze while the remaining system stays in the Coulomb phase.\cite{semoe15} However, in the present PD phase the situation is reversed: The fluctuations of the ordered ``bulk'' state stabilize a disordered phase of a subsystem.
Second, the model for correlated PD discussed here also bears similarities to fractionalized Fermi liquids (FL$^\ast$), in which conventional electronic charge carriers coexist with local moments which itself form a fractionalized spin liquid.\cite{flst1,flst2} One conceptual difference is that the two subsystems of FL$^\ast$ can be adiabatically decoupled in two-band models, while such a decoupling is not possible in the present PD phase.

\subsection{Experimental signatures}\label{subsec:exps}

As the defining feature of PD is the emergence of a conventionally ordered and a quantum disordered subsystem, its qualitative properties can be obtained by combining the ones of the two subsystems.

Assuming that the disordered component is a quantum spin liquid phase with fractionalized quasiparticles, the full system will then feature \emph{both} sharp spin-wave modes associated with the N\'eel-ordered honeycomb component and a continuum of fractionalized quasiparticles, which can be probed by inelastic neutron scattering. The separation of energy scales will lead to a magnon bandwidth of order $J$, while the continuum of fractionalized quasiparticles is on the order of $J'^2/J$. These two components will also show different neutron polarization dependencies.

The fact that 1/3 of the spins of the system are quantum disordered due to frustration can also be observed in thermodynamic signatures. In particular, the entropy per site, $S(T)/N$, can be expected to have a plateau at the value of $1/3 \ln 2$ for temperatures on the order of the effective couplings for the central spins, $T \sim J'^2/J$.
In this context we note that, in the two-dimensional model at hand, spontaneous symmetry breaking is forbidden for any $T>0$ due to Mermin-Wagner's theorem.\cite{mermwag} However, at low but finite temperature the correlation length of the honeycomb subsystem is exponentially large, such that  sharp magnon modes are the relevant excitations in this renormalized classical regime, and our analysis remains valid.

\section{Summary and Outlook}\label{sec:out}

We have provided an effective theory for the correlated partial disorder phase detected in Ref.~\onlinecite{gonz18} in the stuffed honeycomb antiferromagnet, by deriving an effective model for the central disordered spins in the presence of collinear background order.
Our perturbative treatment is controlled in the limit $J' \ll J$ and includes subleading corrections in $1/S$. Within our semiclassical analysis, the effective triangular-lattice XXY model undergoes a transition from an in-plane ferromagnet to an out-of-plane stripe phase, the latter driven by competing Ising interactions, at a critical $S_c \simeq 0.646$, with an intermediate incommensurate phase.
With magnetization corrections growing near the transition, we argue that fluctuations due to the competition of ferromagnetic and stripe orders lead to a quantum disordered ground state for the central spins. Given that $S=0.646$ is close to $S=1/2$, we conjecture that this mechanism drives the correlated partial disorder phase observed numerically.\cite{gonz18,shima18}

Our analysis calls for further numerical investigations. First, it would be of particular interest to investigate the obtained effective triangular-lattice XXZ model beyond the semiclassical techniques employed here. A DMRG study could provide further insights into the nature of the quantum disordered phase anticipated for $S=1/2$. Second, one can modify the behavior of the full stuffed honeycomb-lattice model, by introducing additional interactions between the central spins. Based on our results, we predict that explicit antiferromagnetic interactions of order $(J')^2/J$ either between first-neighbor or second-neighbor sites on the central-spin lattice lead to stripe order on the $C$ sublattice.

On the experimental front, the $\sqrt{3} \times \sqrt{3}$-distortion of triangular lattice leading to a stuffed honeycomb lattice is of significance for modelling basal planes in magnetic perovskite $\mathrm{ABX}_3$ compounds. Notably, $\mathrm{RbFeBr}_3$ shows signs of PD at low temperatures, however due to a large easy-plane anisotropy the local moments on the plane are rather thought to be described by an effective spin-1 XY model.\cite{cope97,atmmh83}
The stuffed honeycomb lattice Heisenberg antiferromagnet has also been studied as a model for the triangular lattice spin-$1/2$ antiferromagnet LiZn$_2$Mo$_3$O$_8$, in which 2/3 of the spins are in a disordered state below a critical temperature.\cite{mcq12} It has been suggested that the local moments on an emergent stuffed honeycomb lattice form a spin liquid which is stabilized by the central spins as magnetic impurities (as opposed to the case discussed in this work, in which the central spins enter a quantum disordered phase). However, in these scenarios a frustrating next-nearest neighbor coupling is crucial for destabilizing the magnetic order of the honeycomb lattice.\cite{flint13,flint18}
We note that alternative explanations for the unusual magnetic behavior of LiZn$_2$Mo$_3$O$_8$ have been put forward.\cite{gang16,gang18}


\acknowledgments

We thank L. Janssen for useful discussions. This research has been supported by the DFG through SFB 1143 (project A01) and the W\"urzburg-Dresden Cluster of Excellence on Complexity and Topology in Quantum Matter -- \textit{ct.qmat} (EXC 2147, project-id 39085490).

\appendix

\section{Derivation of the effective model} \label{sec:deriv_details}

We integrate out the fluctuations of the honeycomb spins in the form of magnons by employing a path integral approach.
Denoting bosonic magnon modes as $a,b$ and the central spins by $S$, the partition function in imaginary time for the full system is given by
\begin{equation}
  \mathcal Z = \int \mathcal D[a,b,S] \eu^{-\mathcal{S}_J -\mathcal{S}_{J'}}.
\end{equation}
Expanding the exponential in $J'$, we obtain
\begin{equation}
  \mathcal Z  = \mathcal{Z}_0 \int \mathcal{D}[S] \eu^{-\mathcal S_\mathrm{eff}},
\end{equation}
with the effective action given by
\begin{equation} \label{eq:seff_exp}
  \mathcal S_\mathrm{eff} = \langle  \mathcal S_{J'} \rangle_J + \frac{1}{2}\left(-{\langle \mathcal S_{J'}^2 \rangle}_J + {\langle \mathcal S_{J'}\rangle}_J^2 \right) + \mathcal{O}\left({J'}^3\right)
\end{equation}
where $\langle \mathcal{O} \rangle_J = 1/ \mathcal{Z}_0 \int \mathcal{D}[a,b] \ \mathcal{O} \eu^{-\mathcal S_J}$ denotes the expectation value with respect to the honeycomb magnon action, and $\mathcal{Z}_0$ the corresponding partition function. The actions noted above are to be taken in imaginary time.

\subsection{$1/S$ expansion}

We find the magnons of the honeycomb system in a $1/S$ expansions by employing the Holstein-Primakoff (HP) representation with bosonic modes $a$,
\begin{align}
  S^+_A &= \sqrt{2 S - n_a} a, \quad S^-_A = a^\dagger \sqrt{2 S - n_a} \\
  S^z_A &= S - n_a,
\end{align}
and similarily for $\vec S_B$, after having picked a local basis by rotating the spins on the $B$-sublattice by $\pi$ around the $x$-axis. After expanding in powers in $1/S$, the Hamiltonian  reads $\mathcal H_{J,h} = \mathcal H^{(0)}_{J,h} + \mathcal H^{(2)}_{J,h} + \mathcal{O}(1/S^0)$, where $\mathcal H^0_{J,h}$ corresponds to the classical ground state energy and $\mathcal H^{(2)}_{J,h}$, which is bilinear in the bosons, is given by \eqref{eq:Hjh_lswt}.
After performing a gauge transformation $a_q \to \eu^{-\iu \phi_q} a_q$ with $\phi_q = \arg f(q)$, the Hamiltonian can be diagonalized by a Bogoliubov transformation
\begin{align}
  a_q &= u_q \alpha_q - v_q \beta_{-q}^\dagger \\
  b_{-q}^\dagger &= - v_q \alpha_q + u_q \beta_{-q}^\dagger,
\end{align}
where $u_q = \cosh \theta_q$ and $v_q = \sinh \theta_q$ with $\tanh(2 \theta_q) = |f(q)|/(3+3h)$, yielding Eq.~\eqref{eq:Hjh_lswt_diag}.
The free magnon Green's functions are then given by\cite{fn:dimless}
\begin{equation}
  G^\alpha_q(\iu \omega) = G^\beta_q(\iu \omega) = J (\iu \omega - \epsilon_q)^{-1}.
\end{equation}

\subsubsection{Self-energy correction to magnon propagator}

The leading-order correction to the magnon propagator due to quartic interactions of the HP bosons, Eq. \eqref{eq:quartic}, can be obtained by normal ordering the magnon operators $\alpha$,$\beta$ after a Bogoliubov transformation, or equivalently by a static mean-field decoupling scheme.
To this end, we note that the non-vanishing boson bilinear expectation values in momentum space are given by
\begin{subequations}
  \begin{align}
    \langle a_q^\dagger a_k \rangle &= \langle b_{-q}^\dagger b_{-k} \rangle =  v_k^2 \delta_{q,k} \\
    \langle a_q b_{-k} \rangle &= \langle a_q^\dagger b_{-k}^\dagger \rangle = - u_k v_k \delta_{q,k}.
  \end{align}
\end{subequations}
Decoupling into the above channels and Bogoliubov-transforming, one obtains the quadratic Hamiltonian
\begin{equation} \label{eq:hf_selfenerg}
  : \mathcal{H}_J^{(0)}: \ = J \sum_q \Sigma^\HF(q) \left(\alpha_q^\dagger \alpha_q + \beta_q^\dagger \beta_q\right)
\end{equation}
with the Hartree-Fock self-energy given by
\begin{align}
  \Sigma^\HF(q) = -\frac{1}{6} &\frac{9(1+h)- |f(q)|^2}{\omega_q} \nonumber \\  &\times\sum_k \left[ \frac{9(1+h)- |f(k)|^2}{\omega_k} - 3 \right],
\end{align}
which coincides with Oguchi's result for bipartite antiferromagnets.\cite{oguchi60}

Having obtained the magnon self-energy, we note that inverting Dyson's equation for the magnon Green's function, $\mathcal G = G + G \Sigma \mathcal{G}$ yields $\mathcal G_q(\iu \omega)^{-1} = J^{-1}(\iu \omega - \epsilon(q) - J \Sigma(\omega,q))$, with the static self energy given by $\Sigma(\omega,q) = \Sigma^\HF_q$.
It should be emphasized that by summing up all 1-particle irreducible contributions to the propagator correction through the use of Dyson's equation, one automatically takes all powers of $1/S$ into consideration, which leads to an inconsistency in our $1/S$ systematics.
To remedy this inconsistency we use the fact that the self-energy is subleading ($\Sigma \sim S^0$ while $\epsilon(q) \sim S$) and expand the interacting Greens function in powers of $1/S$, obtaining
\begin{equation} \label{eq:greenS}
  \mathcal{G}_q(\iu \omega) = G_q(\iu \omega) + G_q(\iu \omega) \Sigma^\HF_q G_q(\iu \omega) + \mathcal{O}(1/S^3).
\end{equation}

\subsubsection{Magnon-central spin interaction}

Expanding $\mathcal H_{J'}$ in $1/S$ yields the Hamiltonian given in Eq. \eqref{eq:h_vertices}, with the vertices defined as
\begin{subequations}
  \begin{align}
    \Gamma^a_q &= f_A(q) \eu^{-\iu \phi_q}, \quad \Gamma^b_q = f_B(q), \\
    \Gamma^{aa}_{q,k} &= -\eu^{-\iu \phi_k} f^*_A(q), \quad \Gamma^{bb}_{q,k} = f^*_B(q), \\
    \Gamma^{3a}_{q;k,p} &= - f_A(q) \eu^{\iu\phi_k +\iu\phi_p-\iu \phi_{k+p-q} },\\
    \Gamma^{3a}_{q;k,p} &= - f_B(q).
  \end{align}
\end{subequations}
The structure factors $f_A$ and $f_B$ for coupling the $A$ and $B$ sublattices to the central spins are given by $f_A(\vec q) = 1+ \eu^{\iu \vec q \cdot (\vec n_2-\vec n_1)} + \eu^{\iu \vec q \cdot \vec n_2}$ and $f_B(q) = 1+ \eu^{-\iu \vec q \cdot \vec n_1} + \eu^{\iu \vec q \cdot (\vec n_2-\vec n_1)}$.
We now rewrite $\mathcal{H}_{J'}$ in terms of the Bogoliubov modes $\alpha_q$, $\beta_q$, yielding
\begin{align} \label{eq:vertices_bogo}
  \frac{\mathcal{H}_{J'}}{J'} &= \sum_q \left\{ \left[ \left( \sqrt{\frac{S}{2}} \Gamma^\alpha_q + \frac{1}{4 \sqrt{2 S}} \Gamma^{3 \alpha}_q\right) \alpha_q \right.\right. \nonumber
  \\&+ \left.\left. \left(\sqrt{\frac{S}{2}}\Gamma^\beta_q + \frac{1}{4 \sqrt{2 S}} \Gamma^{3 \beta}_q \right) \beta_{-q}^\dagger \right] S^-_{-q} + \mathrm{h.c.} \right. \nonumber \\
   &+ \sum_k \left[ \Gamma^{\alpha \alpha}_{q,k} \alpha_{k+q}^\dagger \alpha_k +\Gamma^{\alpha \alpha}_{q,k} \beta_{-(k+q)} \beta_{-k}^\dagger \right. \nonumber \\ &\left. \left. \Gamma^{\alpha \beta}_{q,k} \alpha_{k+q}^\dagger \beta_{-k}^\dagger + \Gamma^{\beta \alpha}_{q,k} \beta_{-(k+q)} \alpha_{k} \right] S^z_q - S_{0}^z \Gamma_{0,q}^{bb} \right\},
\end{align}
where we have dropped in the interest of brevity all arising three-boson terms after normal ordering in anticipation of the Hartree diagram shown in Fig.~\ref{fig:diags}(c): The boson loops will then just give $\alpha$, $\beta$ occupation numbers, which vanish at $T=0$. However, the normal ordering does give rise to corrections to the vertices $\Gamma^\alpha$ and $\Gamma^\beta$ as shown above.
Note that the last summand arises from re-ordering the $b^\dagger b$ term for notational convenience and is cancelled after normal ordering $\beta \beta^\dagger$. For  further considerations this term will be neglected as it does not contribute to any connected diagrams.
The vertices in \eqref{eq:vertices_bogo} are given by
\begin{subequations}
  \begin{align}
    \Gamma^\alpha_q &= \Gamma^a_q u_q - \Gamma^b_q v_q, \quad \Gamma^\beta_q = - \Gamma^a_q v_q + \Gamma^b_q u_q\\
    \Gamma^{3\alpha}_q &= \left(2 \Gamma^{3a}_q u_q - 2 \Gamma^{3b}_q v_q\right) \sum_k v_k^2 \label{eq:3vk} \\
    \Gamma^{3\beta}_q &= \left(-2 \Gamma^{3a}_q v_q + 2 \Gamma^{3b}_q u_q\right) \sum_k u_k^2 \label{eq:3uk} \\
    \Gamma^{\alpha \alpha}_{q,k} &= \Gamma^{aa}_{q,k} u_{k+q} u_k + \Gamma^{bb}_{q,-(k+q)} v_{k+q} v_k \\
    \Gamma^{\beta \beta}_{q,k} &= \Gamma^{aa}_{q,k} v_{k+q} v_k + \Gamma^{bb}_{q,-(k+q)} u_{k+q} u_k \\
    \Gamma^{\alpha \beta}_{q,k} &= -\Gamma^{aa}_{q,k} u_{k+q} v_k - \Gamma^{bb}_{q,-(k+q)} v_{k+q} u_k \\
    \Gamma^{\beta \alpha}_{q,k} &= -\Gamma^{aa}_{q,k} v_{k+q} u_k - \Gamma^{bb}_{q,-(k+q)} u_{k+q} v_k.
\end{align}
\end{subequations}
For the subsequent analysis it is useful to note the $\Gamma^{\beta\alpha}_{-q,k+q} = \Gamma^{\alpha \beta \ast}_{q,k}$ (and $\alpha \leftrightarrow \beta$ analogous).

\subsection{Longitudinal coupling}

We can now compute the connected diagrams that contribute to Eq.~\eqref{eq:seff_exp} at quadratic order. The two \emph{magnon} diagrams relevant for $\XY$ interactions are shown in Figs. \ref{fig:diags}(a) and (b). As explained above, the magnon loop to be computed in the Hartree correction (b) reduces to a renormalized spin-boson vertex at $T=0$, so that the leading and subleading contributions to the transversal coupling are simply given by a single magnon contraction.
With the Green's functions for the (free) magnons $G^\alpha_q(\tau) = - J \langle T_\tau \alpha_q(\tau) \alpha_q^\dagger(0)\rangle$ and similarily for $\beta_q$ one obtains (setting $\tau'=0$ for convenience)
\begin{align}
  \jxy(\tau,q) = \frac{1}{2} &\left\{ \mathcal G_\alpha(\tau,q) \left[ S |\Gamma^\alpha_q|^2 + \frac{1}{4} \Re(\Gamma^\alpha_q \Gamma^{3 \alpha \ast}_q)\right]\nonumber \right. \\
  &\left.+ (\alpha\to\beta, \tau \to -\tau) \vphantom{ \frac{1}{1} } \right\}.
\end{align}
Fourier transforming [with $\mathcal G^\alpha_q(\iu \omega) = \mathcal G^\beta_q(\iu \omega)$ as given in Eq.~\eqref{eq:greenS}] and taking the static limit $\omega = 0$ then yields Eq.~\eqref{eq:jxy_q}.

\subsection{Transversal coupling}

Turning to the longitudinal coupling, we find that at $T=0$, only particle-particle bubbles for the bosons (as shown in Fig.~\ref{fig:diags}) contribute. For the non-vanishing terms, after changing to the frequency domain (with bosonic Matsubara frequencies $\omega, \nu$), we find
\begin{align} \label{eq:jzz_convo}
  \jzz(\iu \nu,q) =& \frac{-1}{2 J \beta} \sum_{\omega} \sum_k \left\{ |\Gamma^{\alpha \beta}_{q,k}|^2 G_{q+k}^\alpha(-\iu \omega -\iu \nu) G^\beta_{-k}(\iu \omega) \right. \nonumber \\ &\left. + |\Gamma^{\beta \alpha}_{q,k}|^2 G^\alpha_k(\iu \omega) G^\beta_{-k-q}(-\iu \omega +\iu \nu) \right\}.
\end{align}
The Matsubara summations can be evaluated with standard methods to yield, at $T=0$ and using inversion symmetry $\epsilon_{-k} = \epsilon_k$,
\begin{equation}
  \frac{1}{J \beta} \sum_\omega G_{k+q}(-\iu \omega) G_{-k}(\iu \omega- \iu \nu) = \frac{J}{\iu \nu + \epsilon_k + \epsilon_{k+q}},
\end{equation}
and analogous for the second term in Eq.~\eqref{eq:jzz_convo}.
Taking the static limit $\nu \to 0$ yields Eq.~\eqref{eq:jzz_q} in the main text.

\section{Gapless limit by $h\to 0$ extrapolation} \label{sec:extrapol}

The expressions for the effective couplings given in Eqs. \eqref{eq:jxy_q} and \eqref{eq:jzz_q} depend on the staggered field $h$. We evaluate the couplings numerically for fixed $h$ on a lattice with $N \times N$ unit cells up to $N=365$ and Fourier-transform to real space according to
\begin{equation}
  j_{ij} = \frac{1}{N} \sum_q j(q) \eu^{\iu \vec q \cdot (\vec r_i - \vec r_j)}.
\end{equation}
Since at any finite $h>0$ the dispersion $\omega(q)$ is gapped and correlations decay exponentially, we perform a finite-size fit of the form $j_{ij}(N,h) = j^\infty_{ij} (h) + A_h \exp[-c_h N]$, with $j(h), A_h$ and $c_h$ as free parameters.

We now discuss the $h \to 0$ extrapolation which involves non-analyticities. As the gap of $\omega(q)$ closes for $h \to 0$, the momentum summations involved in computing $j(h)$ split in a lattice contribution for momenta $|q| \geq \Lambda$, and a continuum contribution for $\lambda < |q| < \Lambda$, where $\lambda$ and $\Lambda$ are the corresponding IR and UV cutoffs. The leading-order behaviour for small $h$ will thus be given by the leading-order in $h$ of the continuum contribution. Note that the gap $h$ acts as an effective IR cutoff, such that we can either evaluate the integral at $\lambda = 0$ and then take $h \to 0$, or equivalently evaluate the momentum space integral at $h = 0$ and then take the limit $\lambda \to 0$.

\subsection{Transversal coupling}

Proceeding, we separate $\jxy(q)$ in Eq. \eqref{eq:jxy_q} into the leading order and subleading terms. For the leading order, using explicit expressions for the Bogoliubov coefficients yields
\begin{equation}
  |\Gamma^\alpha_q|^2 + |\Gamma^\beta_q|^2 = \frac{ 3 (1+h) \left( |\Gamma^a_q|^2 + |\Gamma^b_q|^2 \right) - 2 |f(q)| \Re \left[ \Gamma^a_q \Gamma^{b \ast}_q \right]}{\omega(q)}.
\end{equation}
Expanding the denominator for small $q$ up to quadratic order, the continuum contribution to the coupling is of the form (setting $r=r_i - r_j$)
\begin{equation}
  j(r) \sim \int \frac{\du^2 q}{\left(2 \pi \right)^2} \eu^{\iu q \cdot r} \left( \frac{27 h}{\omega(q)^2} + \frac{9(1-h) q^2}{\omega(q)^2}\right).
\end{equation}
The squared dispersion in the long-wavelength limit reads
\begin{equation} \label{eq:cont_disp}
  \omega(q)^2 \simeq 9h (h+2) + 3/2 q^2.
\end{equation}
Counting powers of momenta, we find that the second term gives a regular contribution, while the first integral formally is $\log$-divergent. This divergence however is cancelled by the factor $h$ originating from expanding the vertex: Explicitly, the first integrand of the form $\eu^{\iu q \cdot r} h/(q^2+h)$ can be evaluated analytically to yield a modified Bessel function (having absorbed numerical prefactors),
\begin{equation}
  j(r) \sim \frac{h}{4 \pi} K_0(h r^2),
\end{equation}
such that no divergent terms appear as $h \to 0$. Expanding the Bessel function, we thus obtain a scaling function
\begin{equation} \label{eq:scaling_leading}
  j^{\XY,0}_{ij} (h) \sim j^{\XY,0}_{ij} + A h \log h + B h + C h^3 \log h + D h^3
\end{equation}
for $h \ll 1$. The prefactors $A, \dots, D$ depend on the distance $(r_i - r_j)$ to the respective neighbor, and can be determined by a non-linear fitting routine.

\begin{figure}[!tb]
\includegraphics[width=.9\columnwidth,clip]{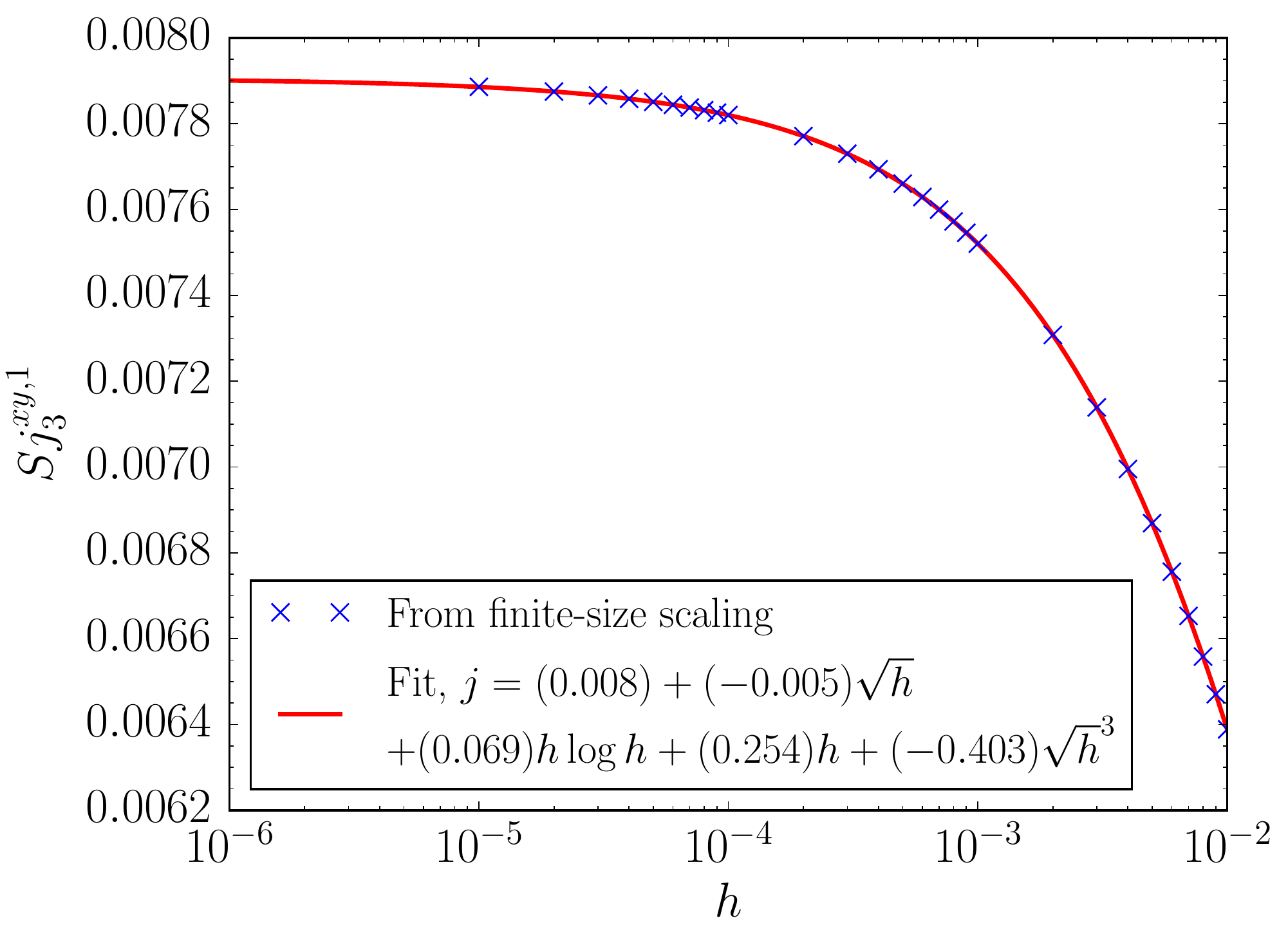}
\caption{
  Example for the $h\to 0$ extrapolation (here: subleading contribution to transversal coupling between third-nearest neighbors) by fitting finite-size scaled data to the ansatz given in Eq.~\eqref{eq:scaling_subleading}. The obtained errors for the fitting parameters are found to be less than $1\%$.
}
\label{fig:example_extrapol}
\end{figure}

The analysis for the subleading terms in $\jxy$ proceeds similarily. Crucially, we note that the vertices $\Gamma^{3 \alpha}_q$ and $\Gamma^{3 \beta}_q$ in Eqs.~\eqref{eq:3vk} and \eqref{eq:3uk} each contribute a constant factor (i.e. independent of $q$) that involves a momentum summation over the Bogoliubov coefficient.
Using that $v_k^2 \sim \cosh(2 \theta_k) = |f(k)| / \omega(k)$ and similarily $u_k^2 \sim \sinh(2 \theta_k) = 3(1+h) / \omega(k)$ we find that the leading-order contribution to the $h\to 0$ scaling due to the summation is of the form (note $\Lambda^2 \gg h$)
\begin{equation}
  \int_{k < |\Lambda|} \du^2 k \frac{1}{\sqrt{k^2+h}} = - \sqrt{h} + \sqrt{h+\Lambda^2} \sim \sqrt{h} + \mathcal{O}(h).
\end{equation}
Since these factors are $q$-independent, they also multiply also a $h$-independent terms appearing in the full evaluation of the subleading corrections, multiplying the scaling form of $j^{\XY,0}$ by a factor $\sqrt{h}$. Furthermore, we note that the self-energy at $h=0$
\begin{equation}
  \frac{9(1+h) - |f(q)|^2}{\omega_q} \xrightarrow{h\to 0} \omega_q.
\end{equation}
The scaling behaviour of the self-energy correction to $\jxy$ thus scales similarily to \eqref{eq:scaling_leading}, such that in total the following scaling ansatz for the subleading corrections is assumed
\begin{equation} \label{eq:scaling_subleading}
  j^{\XY,1}_{ij} (h) \sim \jxy_{ij} + A \ h \log h + B \sqrt{h} + C h + D \sqrt{h}^3.
\end{equation}
An example for the $h\to 0$ extrapolation for the subleading term with fit according to the scaling form given above is shown in Fig.~\ref{fig:example_extrapol}.

\subsection{Longitudinal coupling}

We now discuss the scaling behavior of $\jzz_{ij}$. Since the expression \eqref{eq:jzz_q} involves odd powers of $\omega(q)$, which lead to non-analytic behavior (i.e. the continuum limit and $h\to 0$ do not commute), we consider the case of $h=0$ and work at a finite cutoff $\lambda$ as discussed above. For simplicity we consider the first vertex in \eqref{eq:jzz_q} and expand
\begin{equation}
  | \Gamma^{\alpha \beta}_{q,k}|^2 \omega_k \omega_{q+k} \sim \left(2 k^2 + 2 k \cdot q + q^2 - 2 |k| |k+q| \right).
\end{equation}
Using above expansion and that $\omega(k) \sim |k|$ for $h=0$, the continuum term of the bubble diagram reads after some algebra (again dropping prefactors)
\begin{align} \label{eq:jzz_cont}
  &\jzz(q) \sim \int \du^2 k \frac{| \Gamma^{\alpha \beta}_{q,k}|^2 \omega_k \omega_{q+k}}{|k| |k+q| (|k| + |k+q|)} = \int \du^2 k \frac{1}{|k|} \nonumber \\ &- \int \du^2 k \frac{1}{|k+q|} - \int \du^2 k \frac{4 |k| - 4 |k+q|}{k^2- |k+q|^2},
\end{align}
The first two integrals are seen to be regular from power-counting, as are the subsequent Fourier transformations at a finite $\Lambda$. We note that these regular terms will in general be polynomials in $\lambda$. From Eq.~\eqref{eq:cont_disp} it is seen that the mass dimension of the momenta $[k] = 1/2$ and thus the cutoff scales as $\lambda \sim \sqrt{h}$, such that a generic scaling form will be a polynomial in $\sqrt{h}$.

For the third integral in \eqref{eq:jzz_cont}, which we denote by $I(q)$, it is convenient to invert $k \to -k$ and split the integrand,
\begin{align}
  I(q) = \int \du^2 k \frac{4 |k| - 4 |k+q|}{k^2- |k+q|^2}& =  \int \du^2 k \frac{4 |k|}{k^2- |k-q|^2} \nonumber \\
  &-4 \int \du^2k \frac{|k-q|}{k^2-|k-q|^2}.
\end{align}
Shifting momenta $k\to q-k$ in the second integrand, it is seen that both integrands are equivalent.
The evaluation of the integral follows to a large extent the steps involved in the derivation of the RKKY interaction potential in two dimensions.\cite{kitt68,fiklei74} It is convenient to also directly perform the Fourier transform,
\begin{equation}
  I(r) = \int \du^2 q \int \du^2 k \frac{k}{q} \frac{\eu^{\iu q \cdot r}}{q-2 k \cdot q}.
\end{equation}
Using that $\int_0^{2 \pi} \du \phi (q- 2 k \cos \phi)^{-1} = \theta(q- 2 k) 2 \pi \sqrt{q^2 - 4 k^2}^{-1}$ and performing the angular part of the $q$-integral yields
\begin{equation}
  I(r) \sim \int_\lambda^\Lambda \du k \int_1^\infty \du q' \frac{ J_0(2 k r q') k^2}{\sqrt{(q')^2 -1}},
\end{equation}
where we have substituted $q= 2k q'$. If we neglect the UV cutoff for the inner Integral, the result of the $q$-integration can be given in a closed form
\begin{equation}
  I(r) \sim \int_\lambda^\Lambda \du k k^2 Y_0(kr) J_0(kr),
\end{equation}
where $Y_0$ is a Bessel function of the second kind. By inspecting the integrand it is seen that above integral is regular for a fixed $\Lambda < \infty$. To obtain the $\lambda\to 0$ scaling, we consider $k r \ll 1$ and expand the integrand. Integration is then trivial, and the leading order terms are given by $\lambda^3 \log \lambda$ and $\lambda^3$.
Taking into account the mass-scaling of the cutoff as discussed above thus yields a scaling ansatz of the form
\begin{equation}
  \jzz_{ij}(h) \sim \jzz_{ij} + A \sqrt{h} + B h + C \sqrt{h}^3 + D \sqrt{h}^3 \log h.
\end{equation}

\section{Minimal model for \xyfm{} to stripe transition} \label{sec:toy_stripe}

The competition of \xyfm{} and $\tw$ can be studied by means of a simplified toy model. To this end, we consider a Hamiltonian on the triangular lattice with ferromagnetic nearest-neighbor $\XY$ interactions and competing ferromagnetic-antiferromagnetic Ising $\ZZ$ interactions on nearest and second-nearest neighbor bonds,
\begin{align}
  \mathcal H = - \sum_{\langle ij \rangle} \left[\jxy_1 \left( S^x_i S^x_j + S^y_i S^y_j\right) + \jzz_1 \left(S^z_i S^z_j \right)\right] \nonumber \\+ \sum_{\langle\langle i j \rangle \rangle} \jzz_2 S^z_i S^z_j
\end{align}
with $\jxy_1,\jzz_1,\jzz_2 > 0$. For $\tw$ to be the lowest-energy configuration for the Ising terms, we take $\jzz_2 > \jzz_1$.\cite{kaka74} We now introduce a one-parameter family of ground states a $\vec S = S (0,\sin \varphi,\pm \cos \varphi)^T$, where the positive sign is to be taken on the $A,B$ sublattices and the negative on the $C,D$ sublattices of a $4 \times 1$ unit cell on the triangular lattice.

Minimizing the classical energy as a function of $\varphi$, we find that the two configuration \xyfm{} ($\varphi = \pi /2$) and $\tw$ ($\varphi=0$) are degenerate in energy for
\begin{equation}
  -3 \jxy_1 + \left(\jzz_1 + \jzz_2\right) = 0.
\end{equation}
Taking, for simplicity, $\jxy_1 =1$, $\jzz_1 = \lambda$ and $\jzz_2 = 2\lambda$, with $\lambda \in [0,1]$ being a tuning parameter corresponding to $S$ in the effective model (i.e. fixing the relative strength of longitudinal and transversal interactions), we compute the magnetization corrections for both reference states in LSWT as a function of $\lambda$ (see also Sec.~\ref{sec:eff_mod}).
We find that the degeneracy point is (as in the full model) masked by a mixed phase, with primary ferromagnetic in-plane order and a small incommensurately modulated out-of-plane-component. This incommensurate phase extends from $0.91\lesssim \lambda \lesssim 1.0$.



\begin{thebibliography}{99}
\bibitem{fn:dimless}
Note that we define the self-energy $\Sigma$ and the magnon Green's functions $G$ to be dimensionless, as $J$ is the only relevant energy scale for the honeycomb subsystem.
\bibitem{fn:nosingleion}
We have also performed the stability analysis of Sec.~\ref{sec:eff_mod} with the single-ion anisotropy manually set to zero and find that the \xyfm{}-to-$\tw$ transition is located at $S \simeq 0.57$.
\bibitem{fn:no_jprime}
Note that not all $J'$ couplings are displayed for visual clarity.
\end{thebibliography}
\end{document}